# In Search Of Graphane-A two-dimensional hydrocarbon

Nihar Ranjan Ray[1*], A.K.Srivastava[2] and Rainer Grötzschel[3]

[1]*Saha Institute of Nuclear Physics, 1/AF, Bidhhannagar, Kolkata-700064(India)*

[2]*Electron Microscopy Laboratory, Raja Ramanna Center for Advanced Technology (RRCAT), Indore-452013(India)*

[3]*Institute of Ion Beam Physics and Materials Research, Forschungszentrum Rossendorf P.O.Box 510119, D-01314 Dresden, Germany*

*The characteristic results by different diagnostics on a typical hydrogenated diamond like carbon (HDLC) film synthesized directly in a capacitatively coupled RF (13.56 MHz) discharge are described. The results show some typical charecteric features of Graphane as predicted theoretically. First experimental observation of GRAPHANE/GRAPHANE like structure embedded in the HDLC film is described.*

PACS number(s): 81.05.Zx, 68.55.Nq, 81.05.Uw

*corresponding author: niharranjan.ray@saha.ac.in

The principal aim of this brief report concerns straightforward synthesis of a new Hydrocarbon material which has been theoretically predicted recently[1]. According to this theoretical prediction a new Hydrocarbon compound, called Graphane, is an extended two dimensional covalently bonded hydrocarbon. Its fully dehydrogenated counterpart, Graphene, has been the subject of many recent investigations due to its peculiar transport properties[2-3]. Graphane[1] is a semiconductor and because of its structure and low dimensionality, it provides a fertile playground for fundamental science and technological applications. The typical characteristic features of Graphane, as predicted theoretically[1], can be summarized as below: (i) it has hexagonal network of carbon in the $sp^3$ hybridization with C:H=1 and a *c* lattice constant of 1.6nm (ii) every C-C bond connects carbon atoms with hydrogen at opposite sides of the plane of C-C bond or on the same side of the plane; the former type of conformation of C-C bond is called *chairlike conformer* and the latter one is a *boatlike conformer*; the typical C-C bond length for chairlike conformer is 0.152 nm which is similar to the $sp^3$ bond length of 0.153 nm in diamond and that for boatlike conformer it is 0.156 nm. (iii) the characteristic distinctive feature for chairlike conformer and boatlike conformer is the position of C-H bond stretching mode in the FTIR spectrum: for boatlike conformer the mode occurs at 3026 cm$^{-1}$ and for chairlike conformer the mode occurs at 2919 cm$^{-1}$. (iv) the gravimetric hydrogen capacity of the compound is 6-7 wt%. The two very important attractive properties of this material are: it has a very high volumetric and gravimetric hydrogen density and the electrons in its conduction and valence bands will move in a purely two dimensional topology.

Experimental search for Graphane is being carried out in different laboratories[4-9]. The methodologies consist of mainly three types of experiments viz., hydrogenation of different forms of carbon including nanotubes to see hydrogen adsorption on SWCNT[4-7], hydrogenation of different graphenic surfaces[8] and hydrogenation of different bulk structures of graphite and diamond[8-10]. Graphane material has never been synthesized till date. Direct exposure to $H_2$ does not seem to be correct path to produce graphane because, unlike fluorine, hydrogen does not intercalate graphite[11] due to the higher binding energy of $H_2$ ( ~ 2.4 eV/atom ) compared to $F_2$ (~ 1.5 eV/atom ). Thus, synthesis of Graphane has to be directed through a different path.

Thermal/Electrical decomposition[12-14] of carbon containing gases and hydrogen gases under reduced pressure and electrically biased or temperature controlled substrate was used to grow various thin film of carbon, viz., diamond thin film, diamond like carbon thin film, amorphous carbon thin film etc. According to classification diagram for various hydrogenated carbon films[15], there are still various unknown phases (crystalline, amorphous, nano etc.) of carbon. Whether Graphane phase can also exist in these hydrogenated carbon films produced following the routes of thermal/electrical decomposition of carbon containing gases and hydrogen gas has not been explored. One of the authors (NRR), engaged in the synthesis of hydrogenated diamond like carbon (HDLC) films[16-17] and characterizing the same by various techniques, got the information about Graphane recently[18]. Analyzing the characteristic results by different diagnostics on a typical HDLC film synthesized directly in a capacitatively coupled RF (13.56 MHz) discharge[16] of helium (He) into which methane ($CH_4$) and hydrogen ($H_2$) are admixtured and finding out the presence of Graphane phase, if any, in the film is the motto of the present brief report.

In our works on the synthesis of various HDLC films[16-17] at low substrate temperature (~14$^0$ C), we have made bias pretreatment of mirror polished Si(100) for 15 minutes using hydrogen glow discharge at pressure 0.2mbar and followed by bias (-200volts) enhanced nucleation method (BEN)[19] to prepare the HDLC films. The typical deposition conditions of HDLC film whose characteristic results to be analyzed in this brief report are as below: deposition pressure=0.7mbar, deposition time=30 minutes, dc self bias = -200volts, flow rates: He (1500 sccm), $H_2$ (500 sccm) and $CH_4$ (50 sccm).

Transmission Electron Microscopy (TEM) of the sample: Sample for TEM has been prepared by dimpling and ion milling from the substrate side so as to get electron transparent thin region of the film in large area. Sample was characterized by RRCAT 200kV TEM in both imaging and diffraction mode.

Fig.1a shows the representative diffraction pattern from one of the region. Diffraction pattern shows the single crystal like diffraction. The pattern has been indexed by Carbon with hexagonal crystal structure. The lattice parameters are a= 0.2522nm and c= 1.64743nm. The zone axis of the diffraction pattern is found to be [110]. Fig.1b shows the micrograph of the region from where the diffraction pattern has been taken. It shows the single crystalline nature of the film. The Kekuchi line in the micrograph suggests that the film is essentially single crystal hexagonal structure of Carbon. Typical FTIR spectrum of the sample, as shown in Fig.2, shows presence of C-H bond stretching mode in the hydrocarbon sample. Typical Tauc gap by Ellipsometric measurement of the HDLC sample[17] is 2 eV. Elemental composition of thin films by elastic recoil detection analysis (ERDA)[20] technique: The ERDA measurements were performed at the Rossendorf 5 MV tandem accelerator using 35 MeV Cl$^{+7}$ ions as projectiles. A glancing geometry with an incidence beam angle of 15° to the surface and a detection angle of 31° to the beam direction had been chosen. A Bragg peak ionization chamber (BIC) detects both scattered ions and ejected recoils. Besides of the energy signal this detector provides a signal (bragg-peak height) which allows to identify impinging particles according to its atomic number or nuclear charge, respectively. Simultaneously a silicon detector with an Al stopping foil in front suppressing heavy ions signals detects ejected hydrogen recoils. From the two-dimensional data of the BIC, an example is shown in Fig. 3a, the branches of each recoil element are cut out and projected to the energy axis. Then the depth concentration distributions are calculated according to Spaeth[21] et al. using the stopping data of TRIM95[22] and Rutherford scattering cross sections. For the conversion from the atomic area density scale to a depth scale in length units a layer density of 2.9 g/cm3 was assumed as shown in Fig.3b. The typical experimental result on C: H ratio vs. thickness (nm) is shown in Fig.3c. The obtained depth resolution amounts to about 20 nm for all elements except for hydrogen. Here the energy straggling in the stopper foil gives rise to a lower energy resolution and hence a depth resolution of roughly 50 nm. For all considered elements the detection limit is about 0.1 at%.

Assuming that the C: H atomic ratio in the layers is equal and the measured thickness dependence of this ratio (Fig.3c) can be used to determine the certain amount of hydrogen which is not included in these layers. Both hydrogen of a generally observed thin water film at each surface and possible hydrogen at the substrate prior to the deposition of the C: H layer contribute to excess hydrogen. There is no way to separate this three hydrogen signal contributions directly, but the extrapolation to zero carbon coverage provides a value of about $5 \times 10^{16}$ cm$^{-2}$ for the area density of hydrogen off the layer.

Taking this value into consideration we have evaluated for all samples C: H atomic ratios of $4 \pm 0.3$. Therefore a layer of hexagonal structure of C wherein possibility of formation of one atom of H per C atom in our HDLC sample, a stable chemical compound, can be explored as below: the HDLC[16-17] is made of sp$^2$, sp$^3$ carbons and H atoms; according to XPS results[17], sp$^2$ carbon (284.3eV) content is higher than sp$^3$ carbon (285.3eV) content in the HDLC sample. Now we may assume that hexagonal structure of carbon (Fig.1) is formed out of sp2 carbons first and H atoms convert some sp$^2$ carbons into sp$^3$ carbons in this whole hexagonal structure so that C(sp$^3$) :H=1 i.e. one atom of H per sp$^3$ carbon, may be called as Graphane. Therefore our HDLC film may be described as a hexagonal structure of carbons consisting of sp$^2$ carbons and Graphane phase as defined above. Thus the result C:H= $4 \pm 0.3$ should imply [C(sp$^3$) + C(sp$^2$)]/H = $4 \pm 0.3$. According to experimental result in Fig.3c, C: H ratio never becomes 1 even in the thinnest sample (say, zero thickness), which signifies formation of 100 % Graphane phase is impossible in the present experimental results..

Analyzing the characterizing results based on TEM, ERDA, FTIR, Ellipsometric and XPS measurements, we can summarize the results as below:

The TEM results show single crystal hexagonal structure of Carbon with c lattice constant 1.647 nm. According to ERDA and XPS results our HDLC film may be described as a hexagonal structure of carbons consisting of sp$^2$ carbons and Graphane phase i.e. $C(sp^3)$ :H=1. Typical Gravimetric H capacity: 21.7 at% H i.e. 1.6E17 H atoms $cm^{-2}$. Tauc Gap of 2eV, by Ellipsometric measurement, signifies semi conducting property of our HDLC sample. Thickness[16-17] of the HDLC sample is 120 +/-20nm; length and breadths are several tens of cms; hence the film is a 2D HDLC film.

The comparision of theoretical predictions[1] and the experimental results obtained as described above may be concluded as first experimental observation of GRAPHANE/GRAPHANE like structure embedded in the HDLC film[16-17]. Influence of H atoms on the hexagonal structure of $sp^2$ carbons seems to play an interesting role to discover new structures of carbon for fundamental science and technological applications. Hence synthesis of HDLC samples under various deposition conditions and their proper characterization, measuring physical and chemical properties are required further research.

One of the authors (NRR) wants to thank Jayanth Banavar for informing the news of theoretical prediction of Graphane in PRB (2007) , Jorge O.Sofo for useful discussion and comments about the experimental results, Debajyoti Ray for help to understand the structure of Graphane and Graphene using stick & ball method and special thank to G.S.Lodha (RRCAT)) for  help in doing TEM at RRCAT.  Authors want to thank Abhijit Betal, Dipankar Das, Mahendra Babu, S.S.Sil, A. Bal for their technical help.

[1]Jorge O. Sofo, Ajay S. Chaudhari and Greg D. Barber, Phys. Rev. B **75**, 153401 (2007)

[2]K.S. Novoselov et al., Nature (London) **438**, 197 (2005)

[3]D.A.Abanin, P.A.Lee, and L.S. Levitov, Phys. Rev. Lett. **96**, 176803 (2006)

[4] S.P.Chan, G.Chen, X.G. Gong, and Z.F.Liu, Phys. Rev.Lett. **87,** 205502 (2001)

[5]O. Gulseren, T.Yildirim, and S.Ciraci, Phys. Rev. B **66,** 121401 (2002)

[6] G.Lu, H. Scudder, and N.Kioussis, Phys. Rev. B **72,** 205416 (2003)

[7]G.Chen, X.G.Gong, and C.T.Chan, Phys.Rev. B **72,** 045444 (2005)

[8]D.Stojkovic, P.Zhang, P.E.Lammert and V.H. Crespi, Phys.Rev. B **68,** 195406 (2003)

[9]C.Cab, R.de Coss, G.Oskam, G.Murrieta, and G.Canto, Bull. Am. Phys.Soc. **51,** 888 (2006)

[10]A.B.Anderson, L.N.Kostadinov, and J.C. Angus, Phys. Rev. B **67,** 233402 (2003)

[11]M.S. Dresselhaus and G.Dresselhaus, Adv. Phys. **30**, 139 (1981)

[12] S.Yugo, T.Kanai, T.Kimura and T.Mulo, Appl.Phys. Lett. **58(10),** 1036 (1991)

[13]P.K. Bachmann and W.van Enckevort, Diamond Relt. Mater. **1**, 1021 (1992)

[14] J.Robertson, Material Science & Engineering R **37,** 129-221 (2002)

[15]P.K.Bachmann, In *Ullman's Encyclopaedia of Industrial Chemistry*, **A26**, 720-725 (1996)

[16]N.R.Ray and A.N.S.Iyengar, Proceedings of the Sixth International Conference on Reactive Plasmas and 23[rd] Symposium on Plasma Processing, 24-27 January 2006, Matshushima/Sendai, Japan, 2006 edited by R.Hatakeyama and S.Samukawa(ICRP-6/SPP-23), p.583

[17]A.Singha, A.Ghosh, A.Roy and N.R.Ray, J.Appl.Phys., **100**, 044910(2006)

[18]Jayanth Banavar and Jorge Sofo (private communication)

[19]J.Robertson, J.Gerber, S.Sattel, M.Weiler, K.Jung and H.Ehrhardt, Appl.Phys.Lett. **66**(24), 3287 (1995)

[20]U. Kreissig, R. Gago, M. Vinnichenko, P. Fernandez-Hidalgo, R.J. Martin-Palma, J.M. Martinez-Duart, Nucl. Instr. and Meth. **B 219-220** (2004) 908

[21]C. Spaeth, F. Richter, S. Grigull, U. Kreissig, Nucl. Instr. and Meth. **B 140** (1998) 243]

[22]J.F.Ziegler, J.P.Biersack, and U.Littmark, the Stopping and Range of Ions in Solids, Pergamon Press, New York, 1985

Figure captions:

Fig.1a: Typical diffraction pattern of HDLC film in TEM

Fig.1b: Typical micrograph of HDLC film in TEM

Fig.2:   Typical FTIR spectrum of HDLC film.

Fig.3a:  Typical energy spectra of elements of HDLC film in ERDA measurements.

Fig.3b:  Typical depth profile of elements of HDLC film in ERDA measurements

Fig.3c:  Typical ratio of C & H vs thickness(nm) in the HDLC film

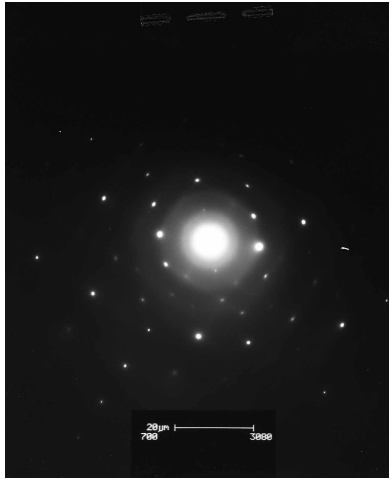

Figure 1a

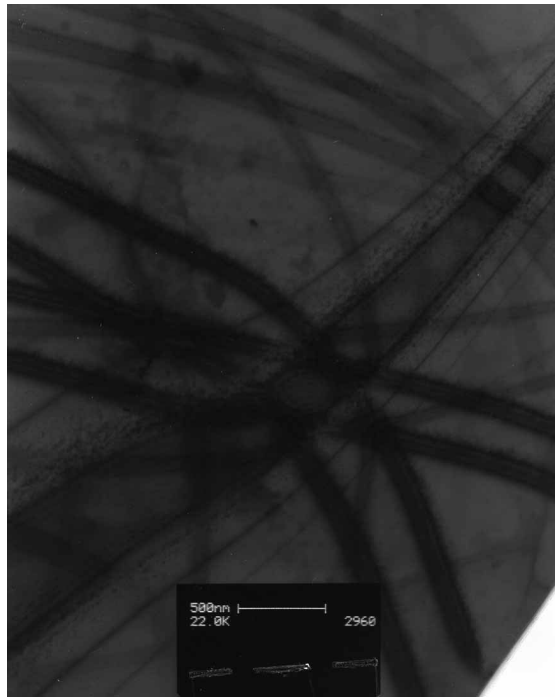

Figure 1b

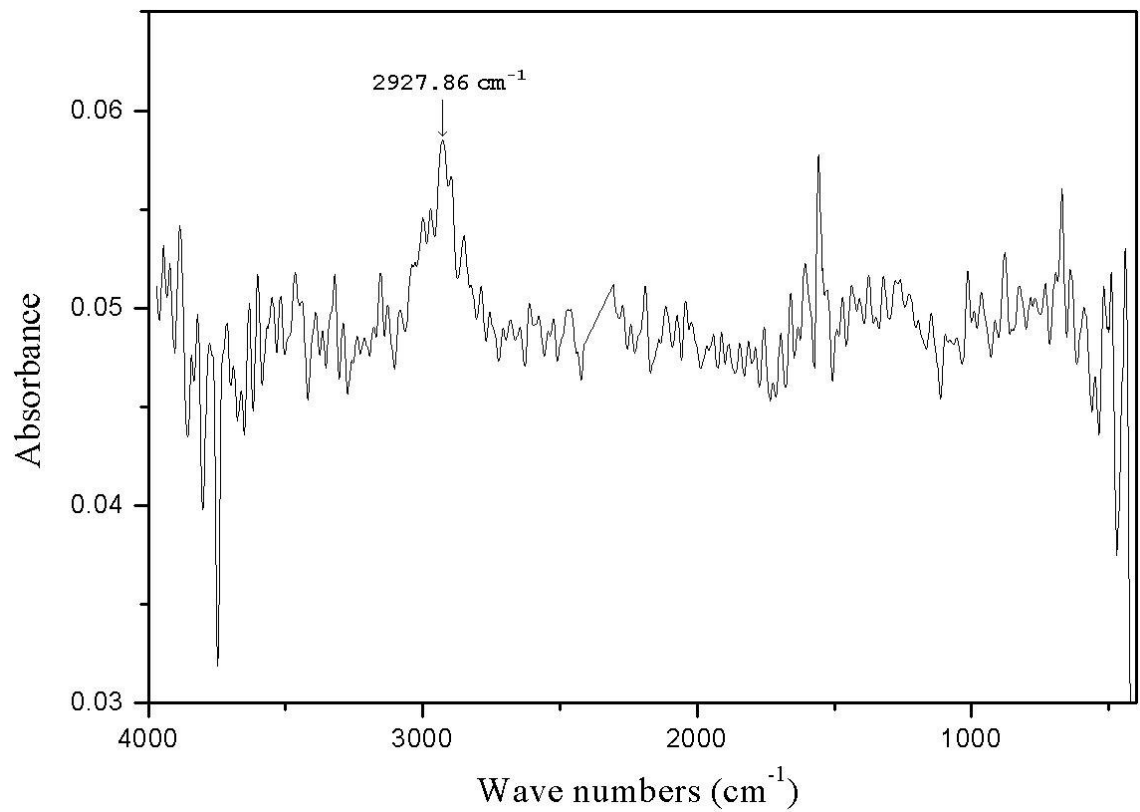

Figure 2

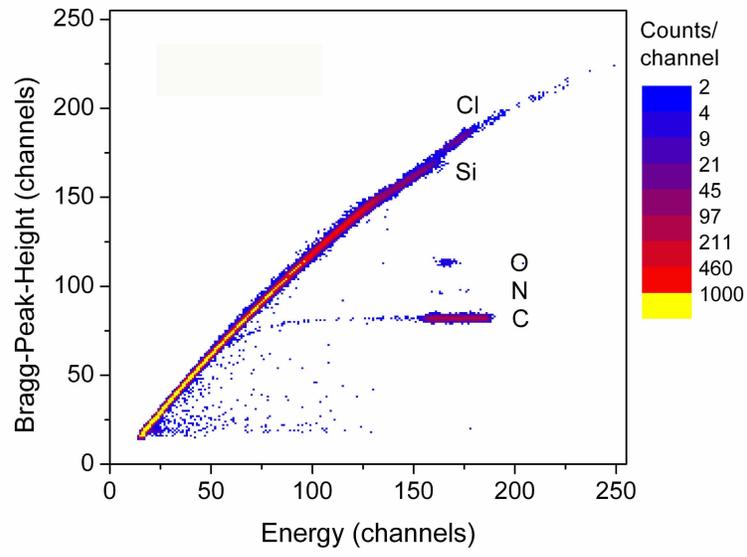

Figure 3a

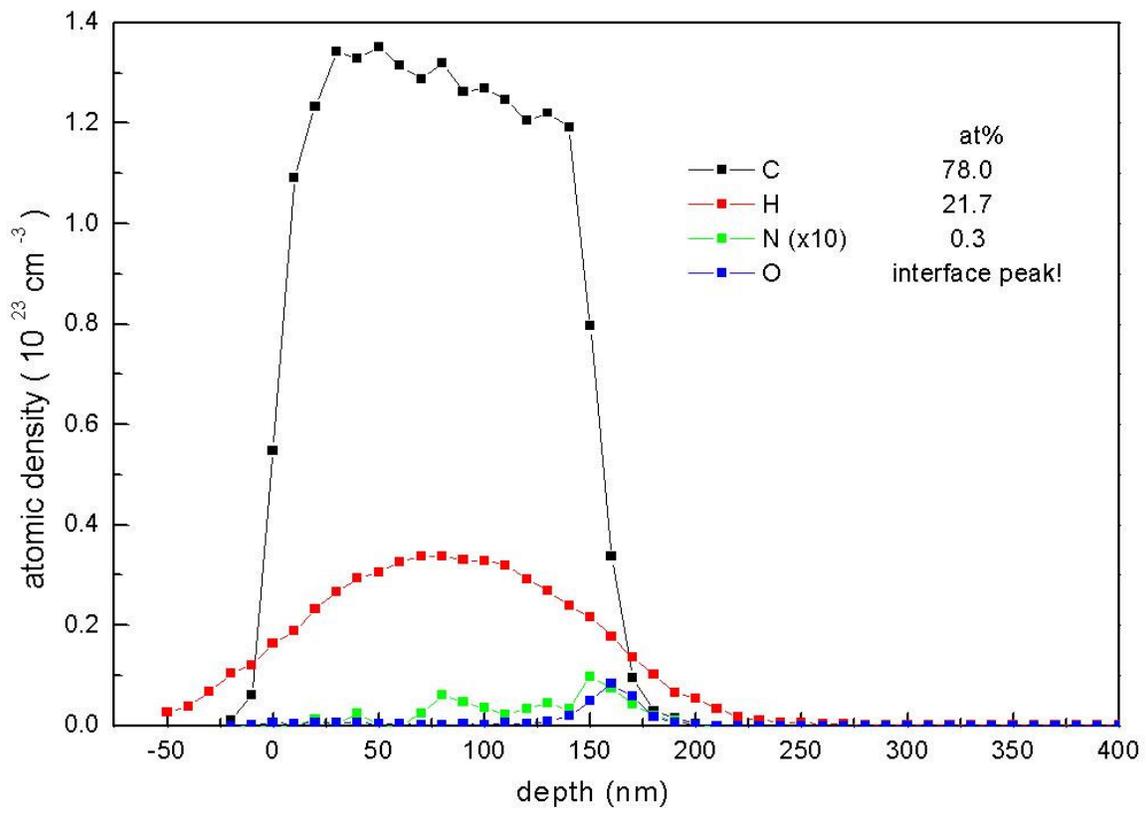

Figure 3b

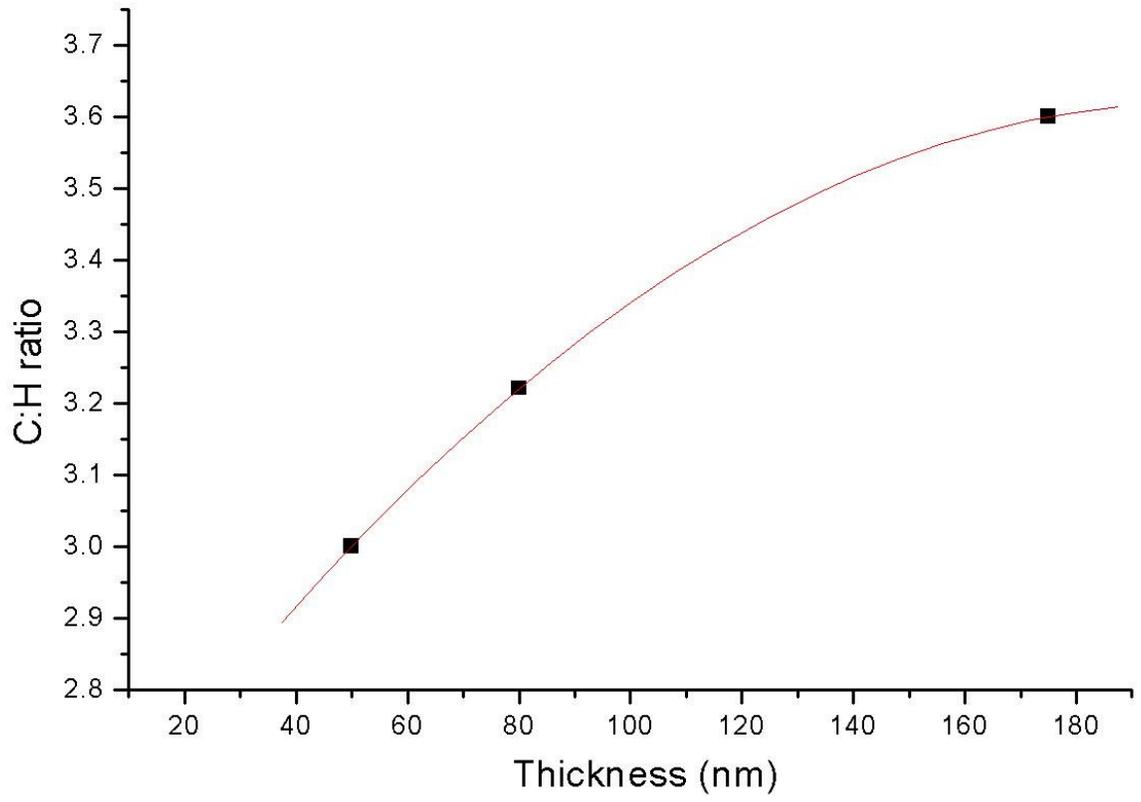

Figure 3c